\documentclass[twocolumn,aps,prc]{revtex4}

\usepackage[english]{babel} 	
\usepackage{graphics} 
\usepackage[pdftex]{graphicx}
\usepackage{amssymb} 			
\usepackage{amsmath} 			
\usepackage{booktabs}
\usepackage{tabularx}
\usepackage{rotating}

\usepackage{float}
\usepackage{url}	
\usepackage{bm}

\graphicspath{{./figures/}}

\allowdisplaybreaks

\begin{document}

\title{Three identical bosons: Properties in non-integer  dimensions and in external fields}

\author{E. Garrido$^{1}$, A.S. Jensen$^{2}$}

\affiliation{$^{1}$Instituto de Estructura de la Materia, IEM-CSIC,
Serrano 123, E-28006 Madrid, Spain}

\affiliation{$^{2}$Department of Physics and Astronomy, Aarhus University, DK-8000 Aarhus C, Denmark}

\date{\today}

\begin{abstract}
Three-body systems that are continuously squeezed from a three-dimensional (3D) space
into a two-dimensional (2D) space are investigated. Such a squeezing can be obtained by means of an external confining potential acting along a single axis. However, this procedure can be numerically demanding, or even undoable, especially for large squeezed scenarios. An alternative is provided by use of the dimension $d$ as a parameter that changes continuously within the range $2\leq d \leq 3$. The simplicity of the $d$-calculations is exploited to investigate the evolution of three-body states after progressive confinement. The case of three
identical spinless bosons with relative $s$-waves in 3D, and a harmonic oscillator squeezing potential  is considered.
We compare results from the two methods and  provide a translation between them, relating dimension, squeezing length, and wave functions from both methods. All calculations are then possible entirely within the simpler $d$-method, but simultaneously providing the equivalent geometry with the external potential.
\end{abstract}






\maketitle

\section{Introduction}             

Specific cold atomic or molecular gases can be controlled by external
fields to previously unprecedented accuracy \cite{blo08,den16}. This
manipulation consists of two relevant ingredients, i.e., (i) the
two-body effective interactions can be varied continuously between
strong attraction and strong repulsion, still while the system is
confined by an external trap, and (ii) the geometric space allowed by
the particles can be designed at will and restricted to
large volumes in three dimensions (3D), flat or curved surfaces (2D),
linear (1D), or anything between these geometries. 
The properties of the systems differ substantially depending
on the confinement,  which in practice can be
varied by use of an external deformed potential, where one or more
dimensions can be squeezed down to vanishing size.  

Together with the dimension, $d$, the properties
depend as well on the number of particles \cite{arm15}.  However, so far only
the $d$-dependence of the relative motion of the simplest systems has
been studied by various methods \cite{yam15,san18,moe19}.  Two
particles squeezed between integer dimensions are obviously the
simplest case, but beside its inherent interest, it is also necessary
in investigations of three particles.  One advantage is that the two
masses only enter in the relative motion as the reduced mass, and only as a
factor in the overall scale parameter.  This is reported in previous
papers using both momentum-space coordinates \cite{san18,ros18} and
ordinary space coordinates \cite{lev14,gar19a,gar19b}.

Recently,
a $d$-dependent formulation  has been presented and applied to two-body systems
\cite{gar19a,gar19b}.  The basic assumption is to use $d$ as a
parameter that can take non-integer values, in such a way that the
external squeezing potential does not appear at all, but is instead
substituted by the correspondingly modified Schr\"{o}dinger equation
depending on $d$ and particle number, $N$ \cite{nie01,jen04}.  The
required numerical effort is similar to a
standard calculation for an integer dimension, but where the external
potential has disappeared. In these works, \cite{gar19a,gar19b}, the
equivalence between the $d$-dependent method and the more
direct procedure working in three dimensions including explicitly
the external squeezing potential was investigated. For the case
of a squeezing harmonic oscillator potential a connection between
the oscillator length and the equivalent dimension $d$ was found.

In this work we extend the method presented in \cite{gar19a,gar19b} to
three-body systems. First, we describe in Section~\ref{sec3} how the
hyperspherical adiabatic expansion method can be implemented to study three-body systems in a $d$-dimensional space. In Section~\ref{sec1} we then
describe how the adiabatic expansion can  be used as well
to treat the same problem in a direct way, i.e., describing the system
in 3D, but introducing explicitly the
external squeezing potential. This procedure, although formally not very
complicated, often leads to calculations that, specially for
large squeezing, are out of numerical reach. The connection between the
dimension and the external field, and the interpretation of the $d$-dimensional wave function
are described in Sections~\ref{secbd} and \ref{d3bwf}, respectively.

The two methods are applied to systems made of three identical
spinless bosons and relative $s$-waves in 3D.
In Section~\ref{sec4} we specify the applied
two-body potentials giving rise to different three-body scenarios.  In
Sections~\ref{sec5} and \ref{sec6} the evolution of the three-body
bound states after progressive squeezing is investigated for each of
the two methods.  A  translation between the parameter, $d$, and the squeezing potential
is investigated in Section~\ref{sec8}, where the wave
function in $d$ dimensions is interpreted as a deformed wave function
in the ordinary 3D-space.  We close in Section~\ref{sec10} with the
summary and the conclusions.

\section{Three bosons in $d$-dimensions}
\label{sec3}

From a general perspective, the description of a given $N$-body system
requires solving the Schr\"{o}dinger equation
\begin{equation} \label{db40}
  \bigg[- \sum_{i=1}^{N} \frac{\hbar^2}{2m_i} \Delta_{\bm{r_i}}
  + \frac{\hbar^2}{2M} \Delta_{\bm{R}_{cm}}+    \sum_{i<j}V_k(\bm{r}_{ij}) - E_{Nb} 
 \bigg] \Psi = 0 \; ,
 \end{equation}
where $m_i$ and $\bm{r}_i$ are mass and position vector of 
 particle $i$, respectively, $M=\sum m_i$ is the total mass, and
$\bm{R}_{cm}$ is the $N$-body center-of-mass coordinate. The
kinetic energy due to the center-of-mass motion is then explicitly removed.
The potential $V_k(\bm{r}_{ij})$ is the interaction between
particles $i$ and $j$, which is assumed to depend on the relative vector, $\bm{r}_{ij}$,
between the two particles. Finally, $E_{Nb}$ is the total $N$-body energy.

The one-body kinetic energy operator for $N$ particles can  be
expressed in terms of  the hyperradius, $\rho$, and all the remaining necessary angles, the hyperangles,
related to the relative degrees of freedom \cite{jen04}.
The square, $\rho^2$, of the hyperradius is defined in terms
of the particle coordinates and the arbitrary normalization mass, $m$, as:
\begin{equation} \label{db50}
\rho^2=\frac{1}{m}\sum_{i=1}^N m_i \left( \bm{r}_i-\bm{R}_{cm}\right)^2
 =  \sum_{i<j} \frac{m_i m_j}{mM} (\bm{r}_i-\bm{r}_j)^2,
\end{equation}
which can be separated into  the Cartesian coordinate contributions, i.e.
\begin{equation} \label{db50b}
\rho^2= \rho_x^2 + \rho_y^2 + \rho_z^2.
\end{equation}

The hyperradial part of the reduced
equation of motion has the usual second derivative operator and a
centrifugal term, $(f-1)(f-3)/(4\rho^2)$, where $f$ is the number of relative 
degrees of freedom.  For a given $N$-body system in a general
$d$-dimensional space it is then clear that $f=d(N-1)$, which results in the
equation of motion in  $d$ dimensions given in \cite{jen04}, i.e.
\begin{eqnarray} \label{db10}
 \bigg[- \frac{\partial^2}{\partial \rho^2} &+&
    \frac{\ell_{d,N}(\ell_{d,N} +1)+ \hat{\Lambda}_{d,N}^2(\Omega_{d,N})   }{\rho^2} + \\ \nonumber
 &+&\frac{2m}{\hbar^2} \sum_{i<j}V_k(\bm{r}_{ij}) - \frac{2m E_{d,N}}{\hbar^2} \bigg] \psi_{d,N}  = 0 \; ,
\end{eqnarray}
where the generalized angular momentum quantum number, $\ell_{d,N}$, is given by
\begin{eqnarray} \label{db30}
  \ell_{d,N} = \frac{f-3}{2}= \frac{1}{2} (d(N-1) - 3),
\end{eqnarray}
and where $\hat{\Lambda}_{d,N}^2$, which depends on the hyperangles $\Omega_{d,N}$,  is the 
generalization to $N$ particles and $d$ dimensions of the usual hypermomentum operator \cite{nie01}.
The energy $E_{Nb}$ in Eq.(\ref{db40}) is denoted now as $E_{d,N}$, making explicit the dependence
on the dimension.
Finally, the phase space reduced wave function, $\psi_{d,N}$,
is expressed in terms of the total wave function $\Psi_{d,N}$ as
\begin{eqnarray} \label{db20}
 \psi_{d,N} =  \rho^{\ell_{d,N}+1} \Psi_{d,N} \; .
\end{eqnarray}

Once the two-body interaction potentials, $V_k$, are
defined, different procedures can be used to reduce Eq.(\ref{db10}) 
to a set of equations depending only on the hyperradius \cite{gar18}. 
The key in all of them consists in expanding the wave function
in a certain basis set that contains the whole dependence on the hyperangles
(for instance the eigenfunctions of the $\hat{\Lambda}_{d,N}^2$-operator),
in such a way that projection of Eq.(\ref{db10}) on the different
basis terms immediately leads to a couple set of differential equations 
for the radial function coefficients.

\subsection{The three-body case}

Being more specific, and focusing on three-body systems, $N=3$, the total three-body wave function
in $d$ dimensions will be obtained in this work by solving the Faddeev equations leading to the Schr\"{o}dinger 
equation (\ref{db10}). In particular, the wave function is  written as:
\begin{equation}
\Psi_d=\frac{\psi_d}{\rho^{\frac{2d-1}{2}}}=
\frac{1}{\rho^{\frac{2d-1}{2}}}\sum_n f_n^{(d)}(\rho) \sum_{i=1}^3\Phi_n^{(d,i)}(\Omega_d),
\label{eq4}
\end{equation}
where $\Psi_d \equiv \Psi_{d,3}$, and 
where the angular functions, $\Phi_n^{(d,i)}$, which form a complete basis set, are the 
eigenfunctions, with eigenvalue $\lambda^{(d)}_n(\rho)$,  of the 
$d$-dependent angular part of  the Faddeev equations (see \cite{nie01} for details). 
 
 Once the angular part has been solved, projection of Eq.(\ref{db10}) on these angular functions leads to the following coupled set 
 of differential equations from which the radial wave functions $f_n^{(d)}$ in the expansion (\ref{eq4})
 can be obtained:
\begin{eqnarray}
\lefteqn{ \hspace*{-15mm}
\left[  
-\frac{\partial^2}{\partial \rho^2}+\frac{1}{\rho^2}
\left(\lambda_n^{(d)}(\rho)+\frac{(2d-3)(2d-1)}{4} \right)-\frac{2mE_d}{\hbar^2}
\right]f_n^{(d)}(\rho)} \nonumber\\ & &
  = \sum_{n'}\left(2P_{nn'}(\rho)\frac{\partial}{\partial \rho} +Q_{nn'}(\rho) \right) f_{n'}^{(d)}(\rho),
\label{eq7}
\end{eqnarray}
where the angular eigenvalues $\lambda_n^{(d)}(\rho)$ enter as effective potentials,
 the explicit form of the coupling terms $P_{nn'}$ and $Q_{nn'}$ can be found in \cite{nie01},
and where $E_d$ is the energy of the three-body system moving in the $d$-dimensional space ($E_d \equiv E_{d,3}$).

The method used is just the hyperspherical adiabatic expansion method,
derived in detail in Ref.\cite{nie01} for any arbitrary dimension $d$.
The generalization of the spherical harmonics to $d$ dimensions can be
found for instance in Appendix B of Ref.\cite{nie01}. They depend on
the $d-1$ angles needed to specify the direction of a given vector coordinate
in $d$ dimensions. This of course makes sense for integer values of
$d$, which leads to an integer number of well-defined angles (for
instance two angles for $d=3$ or one angle for $d=2$).  However, when
$d$ is allowed to take non-integer values, the definition of the
angles and therefore the definition of the spherical harmonics is not
obvious.

To overcome this problem we shall restrict ourselves to $s$-waves, i.e. zero relative 
orbital angular momenta between the particles. In this way the angular dependence of the spherical
harmonics disappears (see Appendix~\ref{app}). 

\section{Harmonic Confinement}
\label{sec1}

The method presented in the previous section appears as an alternative
to the natural way of confining an $N$-particle system, which is to put it under the effect
of an external potential that forces the particles to move in a limited region
of space. Therefore, from the theoretical point of view, the
problem to be solved is just the one given in Eq.(\ref{db40}) applied in 3D,
but where both the interaction between particles, $\sum_{i<j} V_k$, and the trap
potential, $V_{trap}$, have to be included.

An important point to keep in mind
 is that the energy $E_{Nb}$ in Eq.(\ref{db40}) is now the total $N$-body energy, in such 
a way that the energy of the squeezed system requires subtraction of the
(diverging for $d\rightarrow 2$) zero-point energy of an $N$-body system  trapped by the potential $V_{trap}$.

In this work we shall consider an external harmonic oscillator potential acting along the $z$-direction.
The frequency of the potential  will be written as:
\begin{equation}
\omega=\frac{\hbar}{m_\omega b_{ho}^2},
\label{eq1}
\end{equation}
where $m_\omega$ is some arbitrary mass, and $b_{ho}$ will be referred
to as the harmonic oscillator length parameter. Obviously, the smaller 
$b_{ho}$ the more confined the particles, and, eventually, for $b_{ho}=0$ the particles
can move only in a 2D-space. 

The relative trap potential (the center-of-mass part is separated and
omitted) can therefore be written as
\begin{equation}
V_{trap}=\frac{1}{2} \omega^2 \sum_{i=1}^N m_i \left(z_i - Z_{cm}\right)^2,
\label{eqt57}
\end{equation}
where $z_i$ and $Z_{cm}$ are the $z$-components of $\bm{r}_i$ and $\bm{R}_{cm}$, respectively, and
 where $Z_{cm}$, together with the kinetic energy term $\hbar^2\Delta_{\bm{R}_{cm}}/2M$ in
Eq.(\ref{db40}), is introduced to remove the contribution from the center-of-mass motion.

The squeezing potential can be written in a more compact way as
\begin{equation}
V_{trap}=\frac{1}{2}m\omega^2 \rho_z^2,
\label{eqtrap}
\end{equation}
where $m$ is the arbitrary normalization mass introduced in Eq.(\ref{db50}), and $\rho_z^2$ refers to
the $z$-contribution of the square of the generalized hyperradius
vector size defined in Eqs.(\ref{db50}) and (\ref{db50b}).

For this particular case,  and making more specific the discussion above,  the
energy of the squeezed system is given by $E_{ext}=E_{Nb}-E_{ho}$, where
$E_{ho}=(N-1)\hbar\omega/2$ is the zero-point energy in the one
dimensional squeezed oscillator for the $N-1$ relative degrees of
freedom.

\subsection{The three-body case}
\label{sec2a}

Let us focus now on the three-body case, and let us assume
an external harmonic oscillator one-body squeezing potential which, as in
Eqs.(\ref{eqt57}) and (\ref{eqtrap}), acts along the $z$-axis. From the
definition of the $\bm{x}$ and $\bm{y}$ Jacobi
coordinates \cite{nie01}, it is not difficult to see that the trap
potential felt by the three particles can be written as:
\begin{eqnarray}
\lefteqn{
\frac{1}{2}\omega^2\sum_{i=1}^3 m_i r_i^2 \cos^2\theta_i =
\frac{1}{2}m\omega^2x^2\cos^2\theta_x + 
} \label{trap0}\\ && \hspace*{-5mm}
+\frac{1}{2}m\omega^2y^2\cos^2\theta_y+ 
\frac{1}{2} M \omega^2 R_{cm}^2\cos^2\theta_{cm} \nonumber,
\end{eqnarray}
where $\theta_x$ and $\theta_y$ are the polar angles associated to the
$\bm{x}$ and $\bm{y}$ Jacobi coordinates, respectively.

Therefore, after removal of the three-body center of mass motion
the total trap potential takes the form:
\begin{equation}
V_{trap}(x,y,\theta_x,\theta_y)=\frac{1}{2}m\omega^2x^2\cos^2\theta_x + \frac{1}{2}m\omega^2y^2\cos^2\theta_y,
\label{trap2}
\end{equation} 
which is nothing but  the particularization to three particles of Eq.(\ref{eqtrap}).

A more convenient way of writing the trap potential can be obtained by working from
the beginning in the three-body center-of-mass. This means that all coordinates are measured relative to the
center-of-mass, $\bm{R}_{cm}$. In turn, this implies that the
coordinate $\bm{r}_{i}-\bm{R}_{cm}$ corresponding to particle $i$ is
proportional to the $\bm{y}$-Jacobi coordinate in the Jacobi set $i$.
Formally, we can then insert $\bm{R}_{cm}=0$, and in the
center-of-mass system arrive at, see \cite{nie01}
\begin{equation}
\bm{r}_i=\sqrt{\frac{m}{m_i}} \sqrt{\frac{m_j+m_k}{m_i+m_j+m_k }}\bm{y}_i,
\end{equation}
from which the potential in Eq.(\ref{trap0}), or Eq.(\ref{trap2}), can also be written as:
\begin{equation}
V_{trap}= \sum_{i=1}^3 V_{trap}^{(i)}=
\frac{1}{2}m\omega^2 \sum_{i=1}^3 \frac{m_j+m_k}{m_i+m_j+m_k} y_i^2 \cos^2\theta_{y_i}.
\label{trap3}
\end{equation}
This last form of the squeezing potential is particularly useful when,
instead of solving directly the Schr\"{o}dinger equation (\ref{db40}),
the equation is split into its three Faddeev components.

The numerical procedure will be the same as the one shown in the previous section, i.e., we solve the Faddeev
equations in coordinate space by means of the hyperspherical adiabatic expansion method described in \cite{nie01}.
The full calculation is now performed in 3D, and therefore the three-body wave function will be written
as in Eq.(\ref{eq4}) but with $d=3$, that is:
\begin{equation}
\Psi_{ext}=\frac{1}{\rho^{5/2}} \sum_n f_n(\rho) \sum_{i=1}^3 \Phi_n^{(i)}(\rho,\Omega_i),
\label{wf0}
\end{equation}
where $\rho$ is the hyperradius and $\Omega_i$ collects the five hyperangles associated to the Jacobi 
coordinates $\{\bm{x}_i,\bm{y}_i\}$, where $i$ runs over the three possible Jacobi sets \cite{nie01}. 

Again, the angular functions $\Phi_n^{(i)}$ are obtained as the eigenfunctions of the angular part of 
the Faddeev equations, and, as in the previous section, the radial wave functions $f_n(\rho)$ in 
the expansion Eq.(\ref{wf0}) are obtained after solving the coupled set of radial equations (\ref{eq7}),
but now particularized for $d=3$. The energy obtained in this way, $E_{3b}$,  is the total energy, system 
plus external field, in such a way that the energy of the confined system will be  obtained after 
subtraction of the harmonic oscillator energy, i.e., $E_{ext}=E_{3b}-E_{ho}$, which in our case of 
squeezing two coordinates along one direction, see Eq.(\ref{trap2}), means $E_{ext}=E_{3b}-\hbar\omega$.

As shown in \cite{nie01}, the angular functions $\Phi_n^{(i)}$ are obtained after expanding them 
in terms of the hyperspherical harmonics. Calculation of the $\rho$-dependent coefficients in this
expansion requires calculation of the matrix elements of the full potential in between all the hyperspherical harmonics included in the basis set.  Due to the presence of the squeezing term (\ref{trap3}), the calculation of these matrix elements involves calculation of the integral
\begin{eqnarray}
\lefteqn{ \hspace*{-1cm}
W_{\ell_x\ell_y L M}^{\ell'_x\ell'_y L' M'}=
\int d\Omega_x d\Omega_y \left[Y^*_{\ell_x}(\Omega_x)\otimes Y^*_{\ell_y}(\Omega_y)\right]^{LM} } \nonumber \\ &&
\times \cos^2\theta_y
\left[Y_{\ell'_x}(\Omega_x)\otimes Y_{\ell'_y}(\Omega_y)\right]^{L'M'},
\label{wxy}
\end{eqnarray}
where $\ell_x$ and $\ell_y$ are the orbital angular momenta associated to the Jacobi coordinates $\bm{x}$ and
$\bm{y}$, respectively, which couple to the total angular momentum $L$. For simplicity in the notation, 
we have assumed spinless particles, although the generalization to particles with non-zero spin is straightforward.

The integral in Eq.(\ref{wxy}) is analytical, and it takes the form:
\begin{eqnarray}
\lefteqn{ W_{\ell_x\ell_y L M}^{\ell'_x\ell'_y L' M'}=} \nonumber \\ &&
  \delta_{\ell_x \ell'_x}\delta_{M M'}
\sum_{\tilde{L}} (-1)^{\ell_x+M} (2\tilde{L}+1) \hat{L} \hat{L'}\hat{\ell_y}\hat{\ell'_y}
\left( \begin{array}{ccc}
 1  &  1  &  \tilde{L}  \\ 
 0  &  0  &  0
\end{array} \right)^2 
\nonumber \\  && \times
\left( \begin{array}{ccc}
 \ell_y  &  \tilde{L}  &  \ell'_y  \\
 0  &  0  &  0
\end{array} \right)
\left( \begin{array}{ccc}
 L  &  \tilde{L} & L' \\
 -M  &  0  & M
\end{array} \right)
\left\{ \begin{array}{ccc}
 L  &  \tilde{L} &  L'  \\
 \ell'_y &  \ell_x  &  \ell_y
\end{array} \right\},
\label{wxy2}
\end{eqnarray}
where $\hat{\ell}$ means $\sqrt{2\ell+1}$. 

Therefore, the integral Eq.(\ref{wxy}), or, in other words, the trap potential Eq.(\ref{trap3}), mixes the 
relative angular momenta $\ell_y$ and $\ell'_y$, and the total angular momenta $L$ and $L'$,
which is then not a good quantum number (unless the trap potential is
equal to zero).  The projections $M$ and $M'$ are not mixed, and its conserved value determines the 2D 
angular momentum after an infinite squeezing of the particles along the $z$-axis.  

In this work we shall consider the case of three identical spinless bosons, and for the case
of no squeezing only $s$-waves will be considered. In other words, in 3D the system 
will have quantum numbers $L=0$ and $M=0$, which implies that, all along the squeezing
process, the conserved quantum number $M$ will be equal to 0.

\section{Trap versus $d$-parameter}
\label{secbd}

The practical use of the $d$-formalism described in 
Section~\ref{sec3} depends on how to relate to parameters used in the laboratory
setup. A universal connection was established for any two-body system
squeezed from three to two dimensions by an external one-body
oscillator field \cite{gar19a,gar19b}.  The provided relationship then
allows to use the parameter $d$ in the calculation and uniquely relate
to an oscillator frequency or length parameter, or in principle vice versa.

We would like to generalize to three-body systems, but the degrees of
freedom and related structures are now much larger. To begin this
search we start with the very general formulation of $N$ two-body
interacting particles.  As described in the previous two sections, 
the controlling equations are described in
terms of hyperspherical coordinates, where the hyperradius is the most
important coordinate in a widely applicable expression. The reason is
that, if the hyperangles were important, the internal $N$-body structure
would influence the connection.  This in turn can only appear through
a complicated interpretation of the ``spherical'' wave function in the
$d$-calculation.

To understand the relation between $d$ and $b_{ho}$ a little better, we use
a simple model with an oscillator as the internal two-body
interaction. Therefore, the sum of the two-body
interactions entering in Eqs.(\ref{db40}) and (\ref{db10}) takes the form:
\begin{eqnarray} \label{db60}
  \sum_{i<j}V_k(\bm{r}_{ij}) &=& \frac{\omega_{pp}^2}{2M}
  \sum_{i<j} m_i m_j (\bm{r}_{i} - \bm{r}_{j})^2 \; \\ \nonumber
 &=&  \frac{1}{2} m \omega_{pp}^2 \rho^2 \; ,
\end{eqnarray}
where Eq.({\ref{db50}) has been used, and 
where $\omega_{pp}$ describes the strength of the given particle-particle interaction. 

Eqs.(\ref{db40}) and (\ref{db10}) are then pure harmonic oscillator equations, and the 
corresponding energy solutions for the $d$-calculation  and the calculation
with external field, respectively, are \cite{arm15}
\begin{equation} 
  E_d = \hbar \omega_{pp} (\ell_{d,N}+3/2)  = \frac{1}{2} d (N-1) \hbar \omega_{pp}
  \label{db70}
  \end{equation}
  where Eq.(\ref{db30}) has been used,   and
  \begin{eqnarray} \label{db80}
  E_{ext} &=& \hbar \omega_{pp} (N-1) \\ \nonumber   
  &+& \frac{(N-1)}{2} \hbar \sqrt{\omega_{pp}^2 + \omega^2}  
 - \hbar \omega \frac{(N-1)}{2} \; , 
\end{eqnarray}
where the first term is from the two non-squeezed perpendicular
directions, the second term is from the squeezed direction, and the
last term removes the diverging zero-point energy.  The factor $N-1$ on
all terms refer to the number of relative degrees of freedom for $N$
particles.  The two limits of $\omega =0$ and $\omega=\infty$ produce the
oscillator results corresponding to $d=3$ and $d=2$.

If the two procedures are assumed to be equivalent, we then must have
that the expressions in Eqs.(\ref{db70}) and (\ref{db80}) have to be equal,
which after division throughout by $\hbar \omega_{pp}$ results
in
\begin{equation} \label{db90}
 d = 2 + \sqrt{1 + \omega^2/\omega_{pp}^2} - \omega/\omega_{pp} \;, 
 \end{equation}
 and
 \begin{equation} \label{db100}
 \frac{\omega}{\omega_{pp}} = \frac{b_{pp}^2}{b_{ho}^2} = - \frac{(d-1)(d-3)}{2(d-2)} \;,
\end{equation}
where $b_{pp}^2 = \hbar/(m \omega_{pp})$ and $m_\omega$ in Eq.(\ref{eq1}) is taken
$m_\omega=m$.  This relation is independent of the
number of particles, $N$, and it is only an average estimate where all structure
is absent.  This observation is perhaps enhanced by having three
identical bosons in the ground state.

The relation given in Eq.(\ref{db100}) reveals the correct limit at the initial and final
dimensions, that means the squeezing length $b_{ho}$ approaches infinity or
zero, respectively, for the cases of no squeezing ($d=3$) or infinite squeezing ($d=2$).  Also these results 
and conclusions could be achieved from working directly with the special cases of the present
interest $N=2$ and $3$.

When using the harmonic oscillator two-body potentials in Eq.(\ref{db60}), beside the energies Eqs.(\ref{db70})
and (\ref{db80}), the wave functions, solutions of Eqs.(\ref{db40}) and
(\ref{db10}), are also available for the two methods.
 They are very different in structure, since one is
spherical but in $d$ dimensions, and the other is
deformed in three dimensions.  The ground states are Gaussians in both cases
corresponding to oscillators.  

In particular, for the $d$-calculation, the $N$-body ground state oscillator wave function, $\Psi_{N,d}$,
is a Gaussian, which is
\begin{equation}
\Psi_{d,N}\propto 
\exp{\left( -\frac{ \rho^2}{2b_{pp}^2} \right)},
\label{hosol2}
\end{equation}
where $b_{pp}$, as defined below Eq.(\ref{db100}), is the oscillator length associated to the two-body 
interaction (\ref{db60}).

Similarly, for the case of confinement with an external field, the
ground state oscillator wave function, $\Psi_{ext}$, is also a
Gaussian:
\begin{equation}
\Psi_{ext} \propto \exp{\left(-\frac{\rho_\perp^2}{2b_\perp^2}\right)}  
                                      \exp{\left(-\frac{\rho_z^2}{2b_z^2}\right)},
                                     \label{hosol1}
\end{equation}
where the oscillator lengths are
\begin{equation}
b_z^2=\frac{\hbar}{m \sqrt{\omega_{pp}^2+\omega^2}},
\label{hosold1}
\end{equation}
coming from the combination of the two-body and the squeezing
oscillators acting along the $z$-axis, and
\begin{equation}
b_\perp^2=b_{pp}^2=\frac{\hbar}{m\omega_{pp}},
\label{hosold2}
\end{equation}
which results from the two-body oscillator acting on the plane perpendicular
to the $z$-axis,
and where $\rho_\perp$ and $\rho_z$ are defined as:
\begin{equation}
\rho_\perp^2=x_\perp^2+y_\perp^2; \hspace*{1cm}  \rho_z^2=x_z^2+y_z^2,
\label{rhos}
\end{equation}
with $\{x_z, y_z\}$ and $\{x_\perp,y_\perp\}$ being, respectively, the $z$ and perpendicular
components of the Jacobi coordinates $\{\bm{x}, \bm{y}\}$.

Assuming equality between the wave functions in Eqs.(\ref{hosol2}) and
(\ref{hosol1}) from the two methods we infer that the perpendicular
length scale should remain unchanged, while the $z$-direction should
change from $b_{pp}$ to $b_z$.  This results in  a deformation along the $z$-axis of
the $d$-wave function that  reproduces the one of the external field.  The
deformation obviously depends on $d$ or confinement as expressed
through Eq.(\ref{db90}) and Eqs.(\ref{hosold1}) and (\ref{hosold2}).

These ground state wave functions are exact for oscillators, and the
large-distance is asymptotically correct as soon as an oscillator
confining trap is used.  On the other hand, with the $d$-method the
asymptotic wave function for a bound state is falling off
exponentially outside the determining short-range potential.

A more appropriate short-range interaction than the oscillator could
clearly be found, e.g. a gaussian of finite range.  Instead
we shall in the next section elaborate and improve on the above
approximations.

\section{The $d$-dimensional wave function}
\label{d3bwf}

Let us now compare the two three-body wave functions, $\Psi_d$ in
Eq.(\ref{eq4}), and $\Psi_{ext}$ in Eq.(\ref{wf0}), each obtained with
their corresponding methods.  The values of the dimension, $d$, and
the external field parameter, $b_{ho}$, will be such that the ground
state energy is the same in both calculations.

The comparison between the two wave functions will be done following
closely the procedure used in Ref.\cite{gar19b} for two-body
systems. The main idea, as suggested in the previous section when making equal 
Eqs.(\ref{hosol2}) and (\ref{hosol1}), is to interpret the $d$-dimensional wave
function, $\Psi_d$, as a deformed wave function in the ordinary
3D-coordinate space. The Jacobi coordinates, $\bm{x}$ and $\bm{y}$, in $d$ dimensions
are then redefined in the 3D-space as $\tilde{\bm{x}}$ and
$\tilde{\bm{y}}$, with ordinary 3D Cartesian components. However, 
since the squeezing is assumed to take place  along the $z$-axis 
the deformation will take place along that axis, and the $z$-components
of the 3D vectors $\tilde{\bm{x}}$ and $\tilde{\bm{y}}$ will
be deformed.
Both Jacobi coordinates will be deformed in the same way, which
amounts to assume that all the particles feel equally the squeezing. 

More precisely, the modulus of the Jacobi coordinates is redefined as
\begin{eqnarray}
x\rightarrow \tilde{x}=\sqrt{x_\perp^2+(x_z/s)^2}, 
\label{defx}  && \\
y\rightarrow \tilde{y}=\sqrt{y_\perp^2+(y_z/s)^2}, &&
\label{defy}
\end{eqnarray}
from which it is possible to construct the usual hyperspherical coordinates in 3D, i.e., 
$\tilde{\rho}=(\tilde{x}^2+\tilde{y}^2)^{1/2}$, $\tilde{\alpha}=\arctan(\tilde{x}/\tilde{y})$,
and $\tilde{\Omega}_{\tilde{x}}$ and $\tilde{\Omega}_{\tilde{y}}$ which are the polar and azimuthal angles
giving the directions of $\bm{\tilde{x}}$ and $\bm{\tilde{y}}$.

The scale parameter, $s$, controls the deformation of the wave
function, in such a way that for $s=1$ the wave function is not
deformed, and for $s=0$ only $x_z=0$ and $y_z=0$ are possible, and the
system is therefore fully squeezed into a 2D-space.  The introduction
of the parameter, $s$, and the interpretation of $\Psi_d$ as a deformed wave
function in 3D makes it necessary to renormalize the wave function to
$\tilde{\Psi}_d=\Psi_d/C(s)$, where $C(s)$ is given by:
\begin{equation}
C^2(s) =\int d^3xd^3y |\Psi_d(x_\perp,x_z,y_\perp,y_z,s)|^2.
\end{equation}

As in Ref.\cite{gar19b}, the wave functions, $\Psi_{ext}$ and $\tilde{\Psi}_d$, are compared 
through their overlap, i.e.,
\begin{equation}
{\cal O}(s)=\int d^3x d^3y \Psi_{ext}^*(\bm{x},\bm{y}) \tilde{\Psi}_d(x_\perp,x_z,y_\perp,y_z,s),
\label{ovl}
\end{equation}
which can be used as a measure of the accuracy of the scaling
interpretation of the initially spherical $d$-dimensional wave
function.  The scale parameter could temptingly be extracted as the
value of $s$ maximizing ${\cal O}(s)$.

The deformation choice described by Eqs.(\ref{defx}) and (\ref{defy})
is just an estimate of the actual wave function deformation, where the
scale factor is taken to be constant.  Other different choices are
certainly possible, like for instance considering the scale factor,
$s$, a function of $x_z$ and $y_z$.  In any case the overlap described
in Eq.(\ref{ovl}) must be smaller than 1.  As a general rule, the
smaller the squeezing the closer to 1 the overlap, or, in other words,
the better the deformation interpretation as expressed in
Eqs.(\ref{defx}) and (\ref{defy}).

It is worth emphasizing that the overlap must be unity in both limits
of 3D (no squeezing) and 2D (infinite squeezing). In 3D it is obvious, since
 both wave functions are solution
of the same equation, Eq.(\ref{db10}), for $N=3$ and $d=3$. The maximum
overlap between the two solutions will then obviously be equal to 1,
and no deformation of the $d$-wave function will be necessary (i.e.,
$s=1$).  In 2D the wave functions from the two methods must also be
identical, since they are solutions to the same problem in two
dimensions no matter how they are obtained.  The maximum overlap is
then again equal to 1 but now corresponding to $s=0$.
As it will be shown, the problem in this case is the difficulty in reaching this limit
with the external squeezing potential.

The similarity between the two wave functions, $\Psi_{ext}$ and
$\tilde{\Psi}_d$, can be visualized by expanding $\tilde{\Psi}_d$ in
terms of the angular eigenfunctions, $\Phi_n^{(i)}(\rho,\Omega)$, used
to expand $\Psi_{ext}$ in Eq.(\ref{wf0}), i.e.,
\begin{equation}
\tilde{\Psi}_d(x_\perp,x_z,y_\perp,y_z,s)=\frac{1}{\rho^{5/2}} \sum_n \tilde{f}^{(d)}_n(\rho,s) \sum_{i=1}^3 \Phi_n^{(i)}(\rho,\Omega_i),
\label{wf1}
\end{equation}
which, due to the orthogonality of the angular functions, leads to:
\begin{equation}
\tilde{f}_n^{(d)}(\rho,s)=\rho^{5/2} \int d\Omega \Phi_n^*(\rho,\Omega) \tilde{\Psi}_d(x_\perp,x_z,y_\perp,y_z,s),
\label{rad1}
\end{equation}
where $\Phi_n=\sum_i\Phi_n^{(i)}$ and $d\Omega$ is the usual
phase-space associated to the hyperangles for three particles in three
dimensions.  The difference between the
$\Psi_{ext}$ and $\tilde{\Psi}_d$ is then due exclusively to the
different behavior of the radial wave functions $f_n(\rho)$ and
$\tilde{f}_n^{(d)}(\rho)$.

After this interpretation of the $d$-dimensional wave function, based on Eqs.(\ref{defx})
and (\ref{defy}),  we have that the wave function Eq.(\ref{hosol2}), obtained for the
harmonic oscillator two-body interaction Eq.(\ref{db60}), can be understood as an ordinary
3D wave function, which has the large-distance asymptotic behavior:
\begin{equation}
\tilde{\Psi}_{d,N}\propto \exp{\left( -\frac{ \tilde{\rho}^2}{2b_{pp}^2} \right)}=
\exp{\left( -\frac{ \rho_\perp^2}{2b_{pp}^2} \right)} 
\exp{\left( -\frac{\rho_z^2}{2s^2 b_{pp}^2} \right)},
\label{hosol3}
\end{equation}
where $\rho_\perp$ and $\rho_z$ are given in Eq.(\ref{rhos}).

For this wave function to  equal that of Eq.(\ref{hosol1}), obtained
with an external field, and keeping in mind that $b_{pp}=b_\perp$, it is simple to see that
we must have:
\begin{equation}
\frac{1}{s^2} = \frac{b_{pp}^2}{b_z^2}=\sqrt{1+\frac{\omega^2}{\omega_{pp}^2}},
\end{equation}
where Eqs.(\ref{hosold1}) and (\ref{hosold2}) have been used. 

Making now use of Eq.(\ref{db100}), we finally obtain a crude
estimate of the relation between the scale parameter, $s$, the squeezing harmonic oscillator parameter, 
$b_{ho}$, and the dimension $d$:
\begin{equation}
\frac{1}{s^2} = \frac{b_{pp}^2}{b_z^2}=\sqrt{ 1+\frac{b_{pp}^4}{b_{ho}^4} }=
\sqrt{ 1+\left(  \frac{(d-1)(d-3)}{2(d-2)}  \right)^2 }.
\label{sbd}
\end{equation}

This estimate is clearly very simple but at the same time independent
of any details referring to the squeezing process.  Thus it must
necessarily be an approximation when applied to large squeezing
corresponding to distances influenced by the short-range interaction.

\section{Results: Two-body potentials}
\label{sec4}

In this work we shall consider three identical spinless bosons. This is a particularly simple case, since the three terms of the confining 
potential  Eq.(\ref{trap3}) are then identical. This fact has the advantage that with the proper choice of the energy and length units the equations of motion become independent of the mass of the particles. For three different masses the dependence of the squeezing potential on the mass ratios becomes unavoidable. 

In particular, we shall take $m_\omega$ in Eq.(\ref{eq1}) and the
normalization mass, $m$, used to construct the Jacobi coordinates equal
to each other, and equal to the mass of each of the particles. The
length unit will be some length characterizing the boson-boson
potential, which in our case will be the range, $b$, of the short-range
interaction acting between them. The energy unit will be taken equal
to $\hbar^2/mb^2$. When this is done, the dependence on $m$ of
Eqs.(\ref{db40}) and (\ref{db10}), or equivalently, of Eq.(\ref{eq7}),
disappears.

In this work we shall consider two different shapes for the
two-body potentials, a Gaussian potential and a Morse-like potential.
After choosing the range of the interaction as length unit, these two 
potentials can be written as $S e^{-r^2}$ and $S (e^{-2r}-2e^{-r})$, respectively. 
For each shape we have taken two different potentials, whose corresponding
strengths are given in the second column of Table~\ref{tab1}. These
potentials give rise to a series of bound two-body states in 2D and
3D, whose corresponding binding energies, $E_{2D}^{(2bd)}$ and
$E_{3D}^{(2bd)}$, are given in third and fifth columns of the Table,
respectively. In the fourth and sixth columns we give for each
potential the scattering length in 2D, $a_{2D}$, and in 3D, $a_{3D}$,
see ref.\cite{nie01} for definitions.

\begin{table}
\begin{tabular}{|c|ccccc|} \hline
          Pot.       &  $S$  &  $E_{2D}^{(2bd)}$  &  $a_{2D}$  &   $E_{3D}^{(2bd)}$  &  $a_{3D}$ \\ \hline
     A$_g$    &  $-2.86$  &  $-0.538$  &  $1.883$  &  $-3.301\cdot10^{-3} $  &   $18.122$  \\ 
     B$_g$    &$-14.50$  &  $-7.918$  &  $3.916$  &  $-5.075$  &   $-0.201$  \\
                 &                 &   $-0.134$  &                 &          &                  \\ 
     A$_m$    &  $0.95$  &  $-0.148$  &  $3.536$  &  $-3.751\cdot10^{-3} $  &   $18.122$  \\ 
     B$_m$    &$ 3.34$  &  $-1.327$  &  $19.783$  &  $-0.665$  &   $-0.201$  \\
                                          &                 &   $-3.971\cdot 10^{-3}$  &                 &          &                  \\  \hline
\end{tabular}
\caption{ Strengths, $S$, of the Gaussian (A$_g$ and B$_g$)
and Morse (A$_m$ and B$_m$) two-body potentials used in this work. For each of them we give the binding energy of the 
existing two-body bound states in 2D, $E_{2D}^{(2bd)}$, and in 3D, $E_{3D}^{(2bd)}$, as well as the 
corresponding 2D and 3D scattering lengths, $a_{2D}$ and $a_{3D}$. The energies and lengths are given in units 
of $\hbar^2/mb^2$ and $b$,  respectively, where $m$ is the mass of each particle, and $b$ is the range of the interaction.}
\label{tab1}
\end{table}

The Gaussian and Morse potentials will be called potentials A$_g$, B$_g$, and A$_m$, 
B$_m$, respectively. The main difference between them is the number of two-body bound states in two and three dimensions. Potentials A have the same number of bound states, one, in 2D and 3D,
whereas potentials B have one bound state in 3D, but two in 2D. Note that potentials A$_g$ and A$_m$,
and B$_g$ and B$_m$, have the same value of the three-dimensional scattering length $a_{3D}$.
Potentials A have a large value of $a_{3D}$, which is reflected in the small binding of the 3D ground state.
Potentials B have a quite modest and negative value of $a_{3D}$, which indicates that
the first excited state, although unbound, is not very far from the threshold.
Potential B$_m$ has a positive  large value of $a_{2D}$, responsible for the little binding
of the first excited two-body state.
Potentials A are the ones called Potentials II in Ref.{\cite{gar19b}, although in this work the length unit is a 
factor of two smaller than the one used in \cite{gar19b}.

\section{Results: External field case}
\label{sec5}

As mentioned several times already, the most direct way to investigate particle confinement consists 
in including the external potential into the problem to be solved. In our case this means to solve the 
three-body problem as described in Sect.~\ref{sec2a}, where the squeezing potential enters explicitly.

\begin{figure}
\includegraphics[width=1\linewidth]{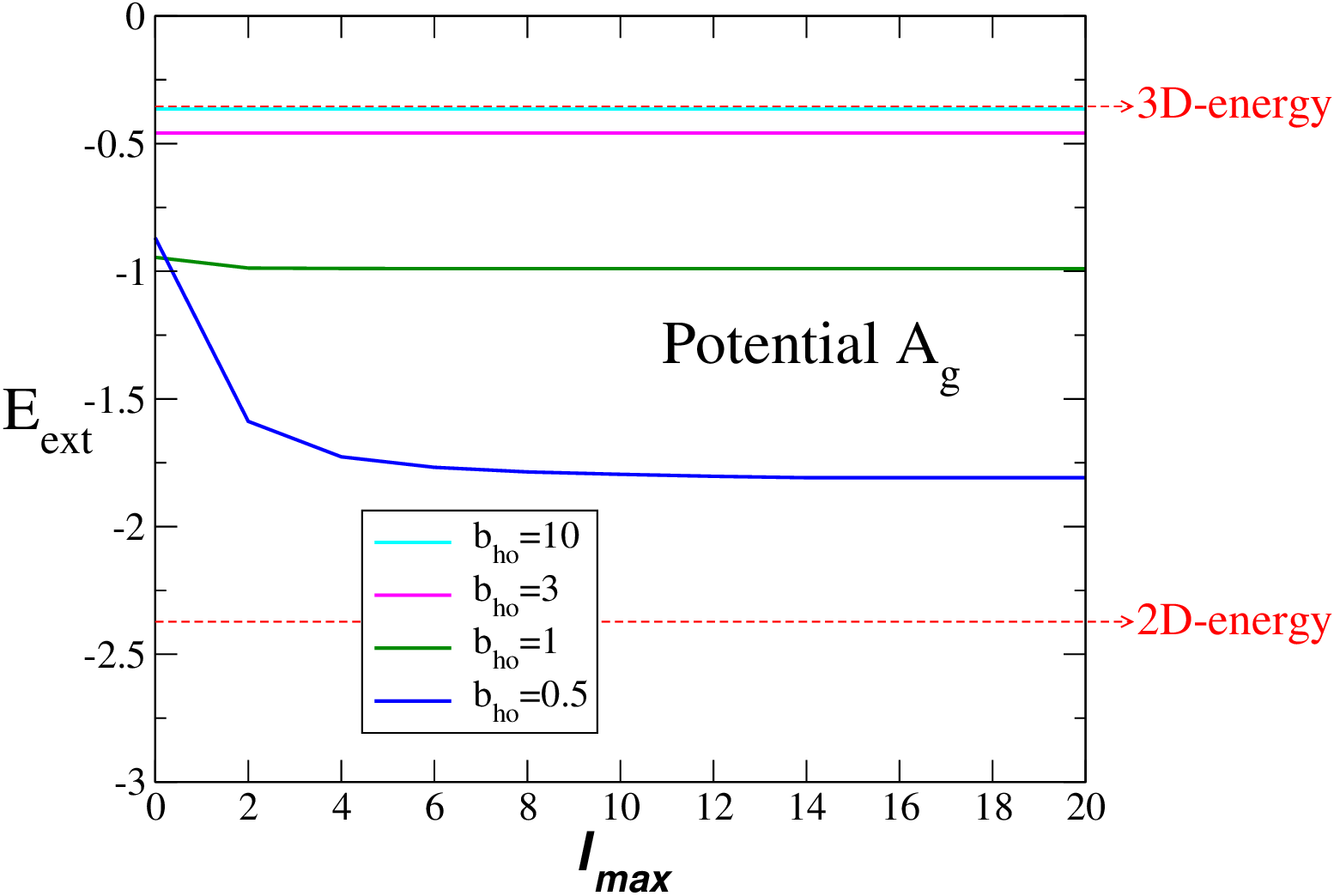}
\caption{For the Gaussian potential A$_g$  and an external squeezing harmonic oscillator field,  we show the three-body ground state binding energy, $E_{ext}=E_{3b}-E_{ho}$, as a
function of the maximum $\ell_y$-value, $\ell_{max}$, used on each Jacobi set.
The results with different values of the squeezing parameter $b_{ho}$
are shown. The lower and higher horizontal dashed lines indicate the 2D and 3D energies, respectively. The energy is in units of $\hbar^2/mb^2$, where
$m$ is the mass of each particle and $b$ is the range of the interaction.}
\label{fig1}
\end{figure}

However, this method presents two main problems. The first one refers to the fact that, as seen in Eq.(\ref{wxy2}), although $\ell_x$ 
(taken equal to zero in this work) is conserved all along the squeezing process, the orbital angular momentum $\ell_y$, and therefore $L$ as well, are not good quantum numbers anymore. This mixing of partial waves is introduced by the squeezing potential, and in fact, the larger the squeezing of the particles the more partial waves are needed. 

This is illustrated in Fig.\ref{fig1}, where we show for the
Gaussian potential A$_g$
in Table~\ref{tab1} how the ground state energy,
$E_{ext}=E_{3b}-E_{ho}$, of the confined three-body system converges
as a function of $\ell_{max}$, where $\ell_{max}$ refers to the
maximum value of $\ell_y=L$ included on each of the Faddeev components.
The lower and higher horizontal dashed lines in the figure indicate
the ground state energy after a pure 2D and a pure 3D calculation,
respectively. The results are shown for different values of the
squeezing parameter $b_{ho}$.

As we can see, for sufficiently large values of $b_{ho}$ the convergence is very fast and just a few $\ell_y$ components, very often only one, are
enough. Eventually, for very large $b_{ho}$ the 3D energy is recovered. With decreasing $b_{ho}$ we observe that higher and higher values of $\ell_{max}$ are needed. 
For instance, for $b_{ho}=0.5$ the components with $\ell_y$ up to at least 14 are necessary to get convergence. For a sufficiently small value of $b_{ho}$ the converged
energy should eventually match the 2D-energy indicated by the lower dashed straight line in the figure. 

\begin{figure}[t]
\includegraphics[width=0.85\linewidth]{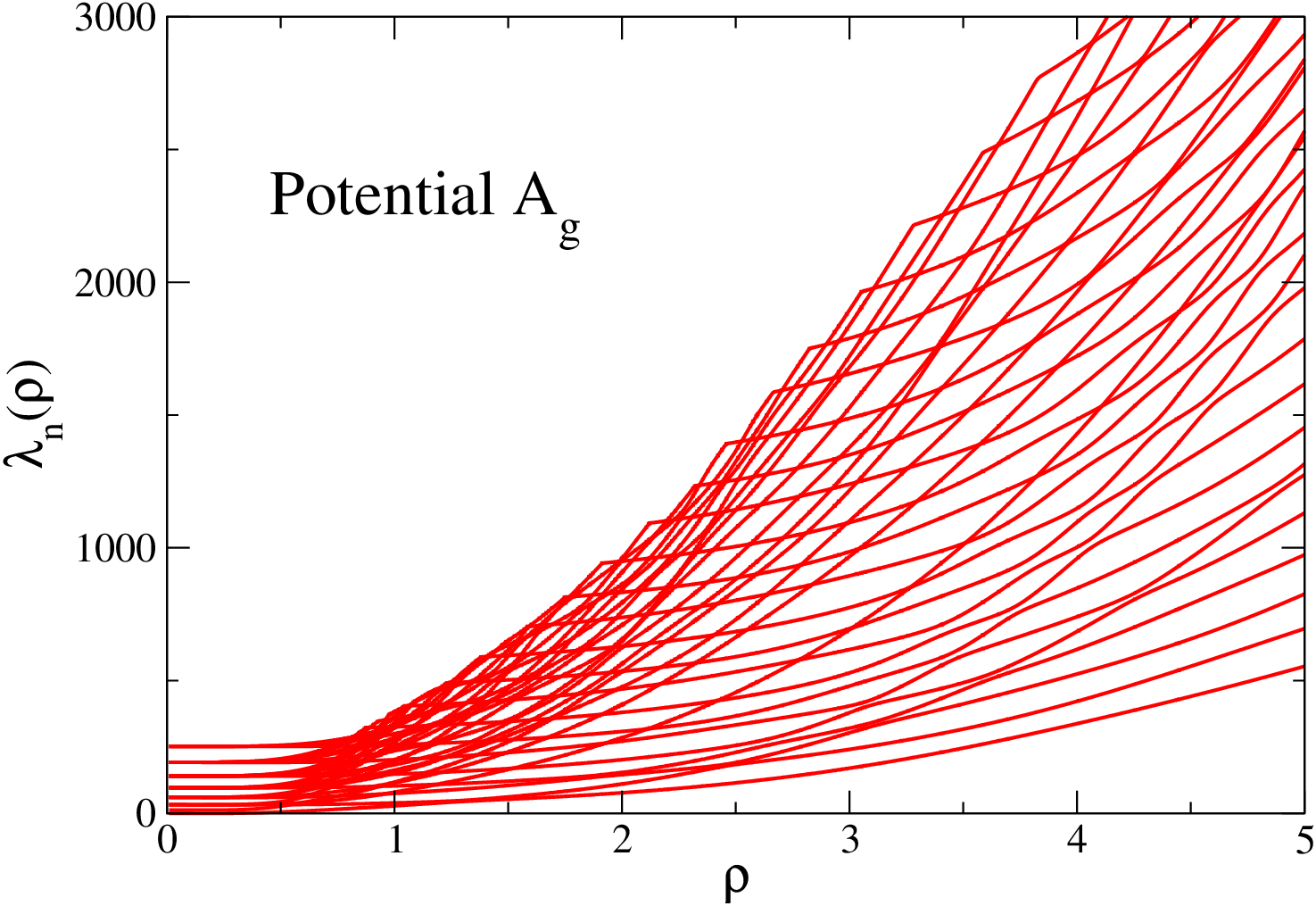}
\caption{Thirty lowest $\lambda_n$-functions for potential A$_g$, squeezing external harmonic oscillator field
with oscillator length $b_{ho}=0.2$, and $\ell_{max}=14$. }
\label{fig2}
\end{figure}

However, to get the same kind of curve for smaller values of $b_{ho}$
is not simple. This is due to the second problem to be faced when
solving the three-body equations with an external potential, which
refers to the convergence of the expansion in Eq.(\ref{wf0}). The
number of terms needed for convergence increases with the confinement
of the particles. This should not be a big problem in itself, except
for the fact that when the squeezing of the particles increases (small
$b_{ho}$-values), the number of crossings between the different
$\lambda_n$-functions entering in Eq.(\ref{eq7}) becomes eventually
too high, which increases dramatically the computing time.  This
problem is actually enhanced by the fact that, as explained above, the
number of required partial waves increases as well with the
confinement.

As an illustration we show in Fig.~\ref{fig2} the 30 lowest
$\lambda_n(\rho)$ functions for potential A$_g$, $b_{ho}=0.2$, and
$\ell_{max}=14$.  To deal with such a huge number of crossings is very
much time consuming, and makes this method rather inefficient close to
2D.  Towards this limit the method of correlated gaussians
\cite{moe19} might for example be used, although the advantage of very
similar basis functions  would  then also disappear.  The choice is then
between using different methods in the two limits or the same method
with inherent loss of efficiency in one of the limits.

\begin{figure}[t]
\includegraphics[width=0.85\linewidth]{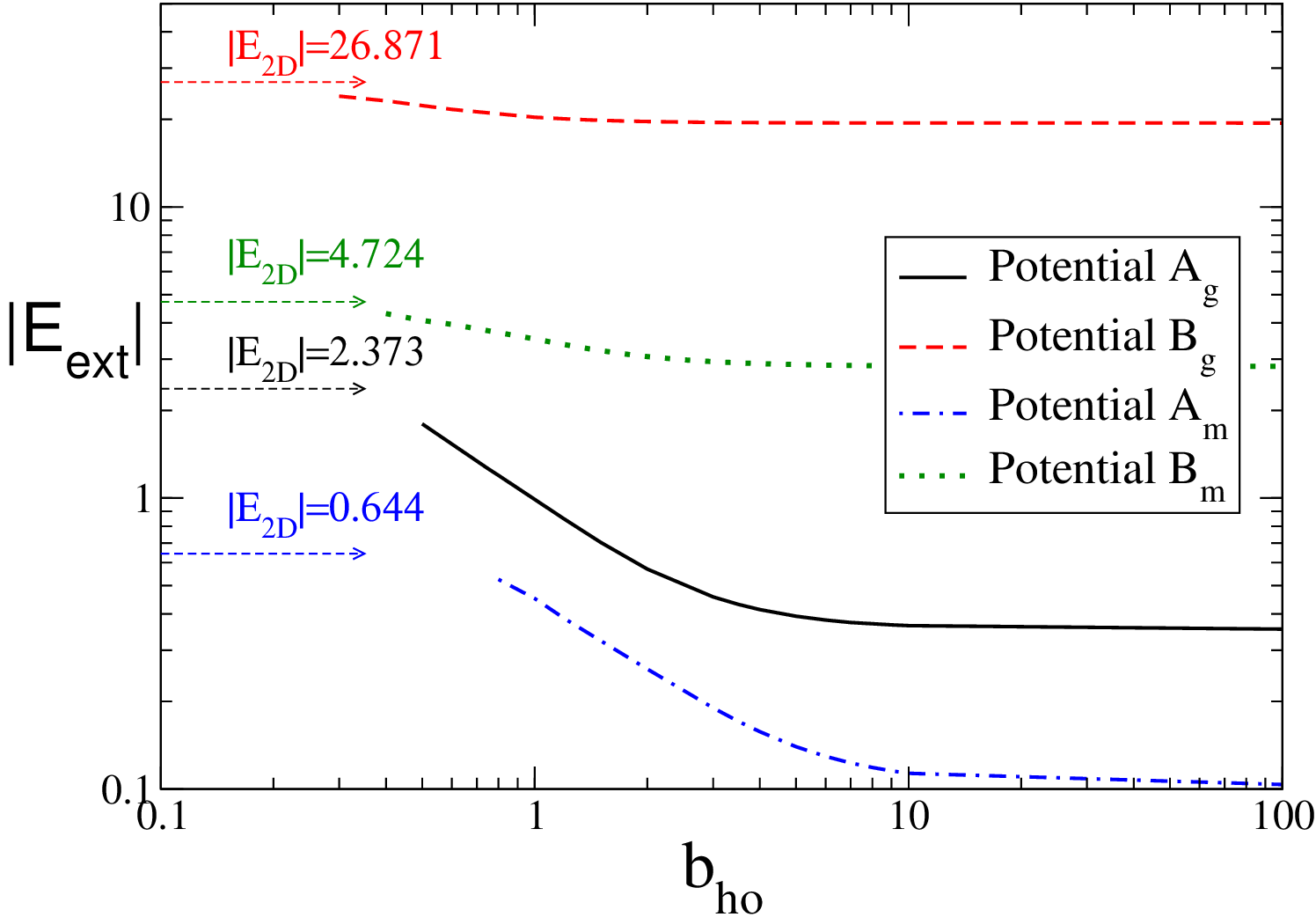}
\caption{For the case of squeezing with an external harmonic oscillator field,
absolute value of the converged three-body ground state energies (in units of $\hbar^2/mb^2$) for 
the Gaussian potentials A$_g$ (solid) and B$_g$  (dashed), and the Morse potentials A$_m$ (dot-dashed) 
and B$_m$ (dot), as a function of $b_{ho}$ (in units of the range of the interaction). The arrows indicate the 
2D energy that should be reached for $b_{ho}=0$.}
\label{fig3}
\end{figure}

The conclusion is therefore that to include explicitly the external
potential as described in Section~\ref{sec2a} is not very convenient
in the case of large squeezing. To reach convergence for low values of
the squeezing parameter becomes at some point too troublesome. In
Fig.\ref{fig3} we show the computed converged energies of the
three-body ground state for the four potentials given in
Table~\ref{tab1} as a function of $b_{ho}$. In the limit of $b_{ho}=0$
the 2D energies indicated by the arrows in the figure should be
reached.

\section{Results: The $d$-calculation}
\label{sec6}

\begin{table}
  \begin{tabular}{|c|cc|cc|}
\hline     
     Pot.         &   $E_{2D}$   & $r_{2D}$ &  $E_{3D}$  & $r_{3D}$ \\ \hline
   A$_g$  &    $-2.373$    &                 0.591                        &  $ -0.354$  &                1.081                                  \\
              &    $-0.556$    &                 3.339         & $ -6.109\cdot10^{-3}$  &               10.280    \\ \hline
  B$_g$  &     $ -26.871$  &            0.304             &       $-19.426$             &           0.390                                                 \\
              &    $-14.622$     &      0.518                  &    $-9.050$                &             0.632                             \\
              &     $   -7.990$    &       1.678               &                                   &                                                    \\ \hline
  A$_m$ &      $  -0.644$      &        1.145             &      $-0.104$             &          2.201                                         \\
                          &       $ -0.160$       &      4.695              &     $-4.808\cdot 10^{-3}$      &        15.744                        \\ \hline
  B$_m$ &      $  -4.724$      &       0.613             &      $-2.834$             &          0.848                                         \\
                          &       $ -2.202$      &      1.220            &          $-1.087$             &           1.628                                            \\ \hline
\end{tabular}
\caption{For $d=2$ and $d=3$, we give the three-body energies and root-mean square  radii, 
$r=\langle \rho^2/3\rangle^{1/2}$, for all the bound states 
obtained with the potentials given in Table~\ref{tab1}. The subscripts $2D$ and $3D$ refer to the results obtained for $d=2$ and $d=3$, respectively.
As done all along the text, the energies are given in units of $\hbar^2/mb^2$, where $m$ is the mass of each of the particles (and the normalization mass used to construct
the Jacobi coordinates), and $b$ is the range of the two-body interaction, which is taken as length unit.}
\label{tab2}
\end{table}

To perform the three-body calculations as described in Section~\ref{sec3}, where the external potential does not enter and the dimension $d$ is treated as a parameter, is certainly much simpler than the calculations shown in the previous section. In Table~\ref{tab2} we give the computed energies and root-mean-square (rms) radii 
in two and three dimensions for all the three-body bound states obtained after solving the coupled Eqs.(\ref{eq7}) with the two-body potentials given in Table~\ref{tab1}. The energies, $E_{2D}$ and $E_{3D}$, and the rms values, $r_{2D}$ and $r_{3D}$, refer to the results obtained with $d=2$ and $d=3$, respectively.
As we can see, whereas potentials A$_g$, A$_m$, and B$_m$ have the same number, two, of bound three-body states in 3D as in 2D, 
potential B$_g$ has one bound state less in 3D, two, than in 2D, three.

In Fig.~\ref{fig4} we show the absolute value of the bound three-body energies obtained with potentials A$_g$, B$_g$, A$_m$, and B$_m$  (panels (a), (b), (c), and (d), respectively) as a function of
$(d-2)/(3-d)$, where $d$ is the dimension running within the range $2\leq d \leq 3$. The choice of the abscissa coordinate as  $(d-2)/(3-d)$ has been made to facilitate the
comparison with Fig.~\ref{fig3}, since in both figures values of the abscissa coordinate equal to zero and infinity correspond, respectively, to maximum squeezing of the system into 2D and
no squeezing (3D). In the figure the dotted (blue) curve shows, for each of the potentials and also as a function of $(d-2)/(3-d)$, the absolute value of the lowest two-body bound state energy.

\begin{figure}[t]
\includegraphics[width=1\linewidth]{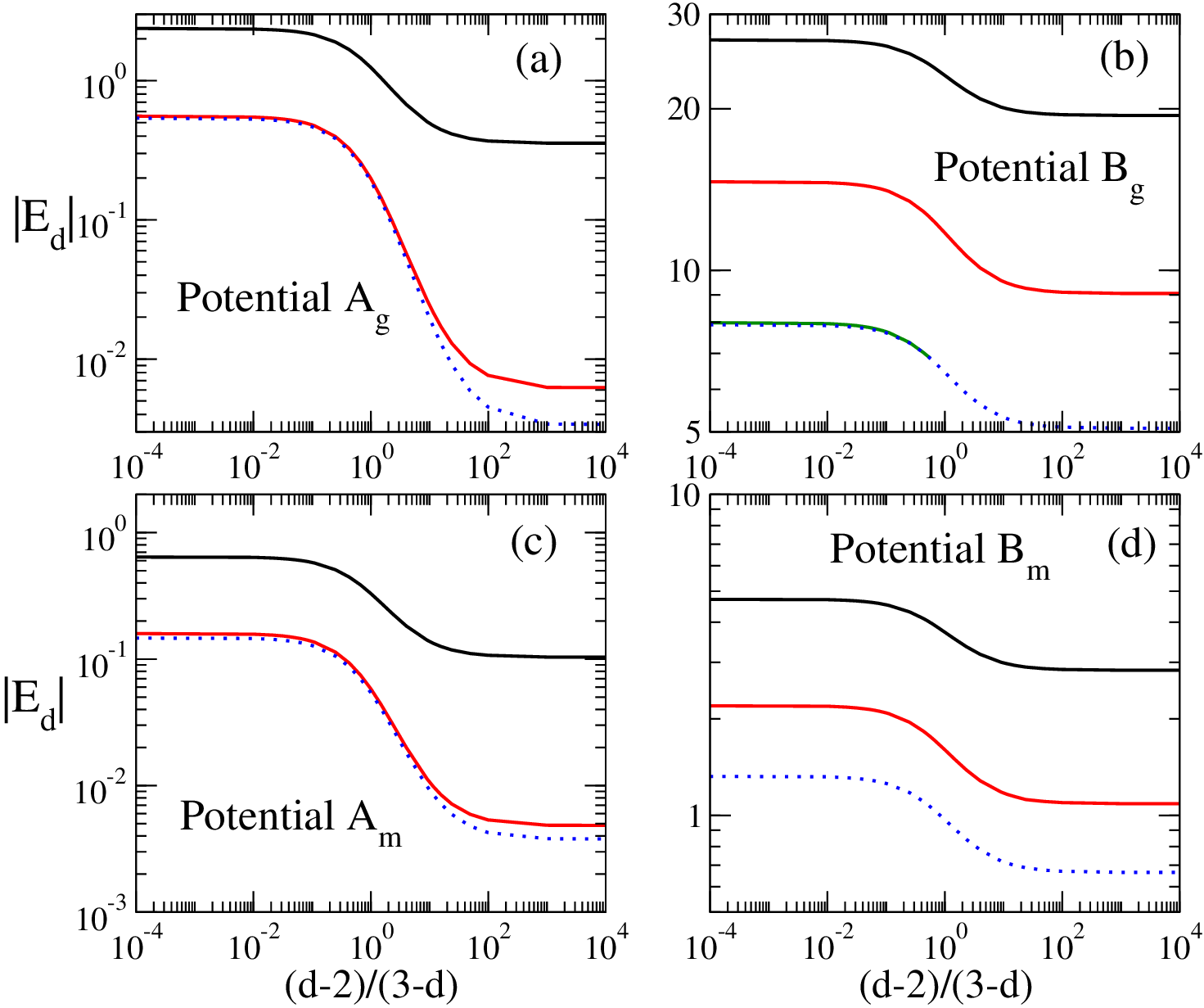}
\caption{With the $d$-method, absolute value
of the bound state three-body energies obtained with potentials A$_g$, B$_g$, A$_m$, and B$_m$
 in Table~\ref{tab1} (panels (a), (b), (c), and (d), respectively) as a function of
$(d-2)/(3-d)$ where $d$ runs within the range $2\leq d \leq 3$. For each of the potentials, the dotted blue curve is the absolute 
value of the lowest bound two-body state energy. }
\label{fig4}
\end{figure}

As a general rule, when squeezing from $d=3$ to $d=2$ the three-body system becomes progressively more and more bound. This is clearly seen for all the potentials when
moving from the right part on each panel ($d=3$) to the left part ($d=2$). 
This is a reflection of the lower centrifugal barrier in 2D than in 3D.

In the case of potentials A$_g$ and A$_m$, Figs.~\ref{fig4}a and \ref{fig4}c, the curve corresponding to the excited three-body bound state (red curve) 
approaches the dotted curve representing the bound two-body
energy. In fact, very soon the three-body excited state is just slightly more bound than the two-body state, representing therefore a boson very weakly bound with
respect to the two-body bound state.

For potential B$_g$, Fig.~\ref{fig4}b, a new bound state shows up for $d\approx 2.4$. It appears from the threshold corresponding to the bound 
two-body state and the third boson. From that point and up to $d=2$ the third three-body bound state follows closely the 
two-body binding energy (dotted curve). Thus, as for potentials A, the last excited state corresponds to a very weakly bound boson with respect 
to the two-body bound state.

For potential B$_m$ the situation seems to be different compared to the other potentials, since in this case there is no three-body bound state approaching
the two-body bound state curve (dotted curve). However, the behavior of the three-body spectrum for this potential is actually very similar to 
the one of potential B$_g$ in Fig.~\ref{fig4}b. The only difference is that the dimension at which the third bound state appears
from the two-body threshold is $d\approx 1.9$, lying therefore out of the graph limits. It would certainly be possible to extend the investigation to
confinement scenarios up to dimensions smaller than 2, eventually up to $d=1$ or even smaller, but this is left for a future work.

\section{Wave functions}
\label{sec8}

In Section~\ref{secbd} the relation between the harmonic oscillator parameter, $b_{ho}$, and the 
dimension, $d$, was discussed. In particular, the relation in Eq.(\ref{db100}) was found for the simple 
case where the particle-particle interaction is assumed to be a harmonic oscillator potential with
length parameter $b_{pp}$. This length, $b_{pp}$, is also the root-mean-square radius of the three-body system
in 3D, whereas in 2D the root-mean-square radius is $r_{2D}=b_{pp}\sqrt{2/3}$, which permits to 
rewrite Eq.(\ref{db100}) as:
\begin{equation}
\frac{b_{ho}}{r_{2D}} = \sqrt{ \frac{3(d-2)}{(d-1)(3-d)} }.
\label{bdrel}
 \end{equation}

Instead of this simple expression, the full numerical solution from the Gaussian and Morse 
short-range interactions can be found, and the energies compared. In the same way that Eq.(\ref{db100}) has been obtained
by making equal $E_d$ and $E_{ext}$ in Eqs.(\ref{db70}) and (\ref{db80}),
we can use Fig.~\ref{fig3} and Fig.~\ref{fig4} to obtain numerically the relation
between $d$ and $b_{ho}$, at least for the $b_{ho}$-values for which,
as shown in Fig.~\ref{fig3}, the energy $E_{ext}$ can be obtained.

\begin{figure}[t]
\includegraphics[width=0.85\linewidth]{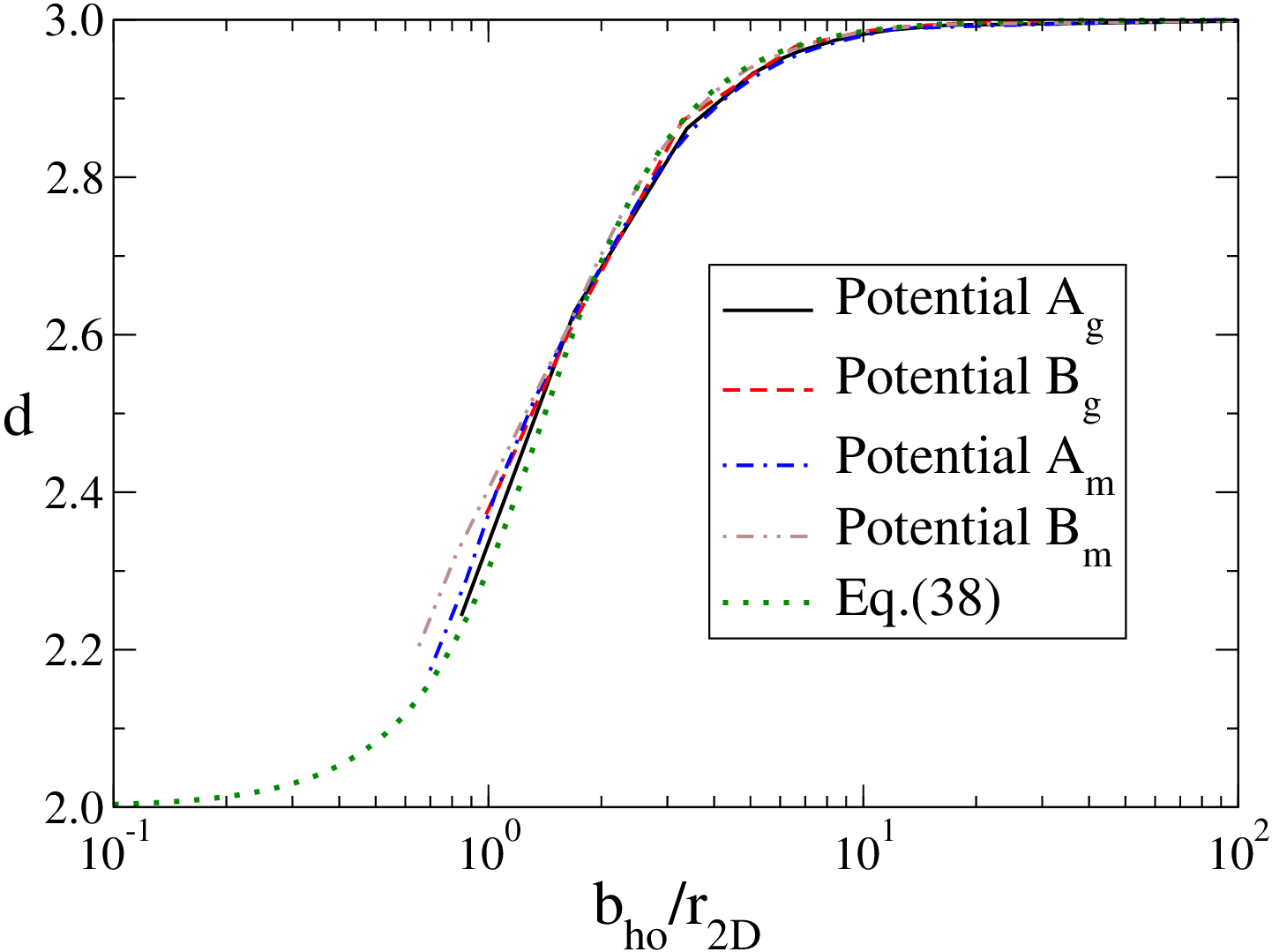}
\caption{Numerical relation between the dimension $d$ and the harmonic oscillator
parameter $b_{ho}$ obtained after making equal $E_{ext}$ in Fig.~\ref{fig3} and the
ground state energy $E_d$ in Fig.~\ref{fig4}. The oscillator parameter $b_{ho}$ is normalized
to the root-mean-square radius of the 2D three-body calculation. The cases of potentials A$_g$, B$_g$, 
A$_m$, and B$_m$ are shown by the solid, dashed, dot-dashed, and dot-dot-dashed curves ,respectively. 
These results are compared with the estimate given in Eq.(\ref{bdrel}), which is shown by the dotted curve.}
\label{fig5}
\end{figure}

The results are shown in Fig.~\ref{fig5} by the solid, dashed, dot-dashed, and dot-dot-dashed 
curves for potentials A$_g$, B$_g$, A$_m$, and B$_m$, respectively. They have been obtained using the energies
of the three-body ground state. The numerical curves are compared with the
analytical expression given by Eq.(\ref{bdrel}), which is shown by the dotted curve.

As we can see, the estimate in Eq.(\ref{bdrel}) agrees reasonably well with the calculations
in those regions where the numerical calculation with the external potential has been possible.
This is especially true for potentials A$_g$, B$_g$, and A$_m$, whereas for potential B$_m$
 a somewhat bigger discrepancy is observed for large squeezing. In any case, for those small values
of the squeezing parameter, $b_{ho}$, the complications inherent to the numerical calculations 
with the external potential can be a source of inaccuracy for $E_{ext}$. Furthermore, since $b_{ho}=0$
must necessarily correspond to $d=2$, we can consider the expression (\ref{bdrel}) a reliable translation to the parameter, $d$, in relatively easy calculations from a given small value of the squeezing parameter, $b_{ho}$.

As shown in Section~\ref{d3bwf}, the equivalence between the three-body wave functions
with the two calculations can be directly compared through the overlap in Eq.(\ref{ovl}), where
$\Psi_{ext}$ is the wave function obtained with the external squeezing potential, and 
$\tilde{\Psi}_d$ is the one obtained with the $d$ calculation after the transformation 
in Eqs.(\ref{defx}) and (\ref{defy}), and the subsequent renormalization. Obviously,
the values of $b_{ho}$ and $d$ have to be such that they both give rise to the same ground state
three-body energy. As discussed in Section~\ref{d3bwf}, this overlap permits to extract
the scale parameter, $s$, in the transformation in Eqs.(\ref{defx}) and (\ref{defy}) as the
value that maximizes the overlap.

\begin{figure}[t]
\includegraphics[width=0.95\linewidth]{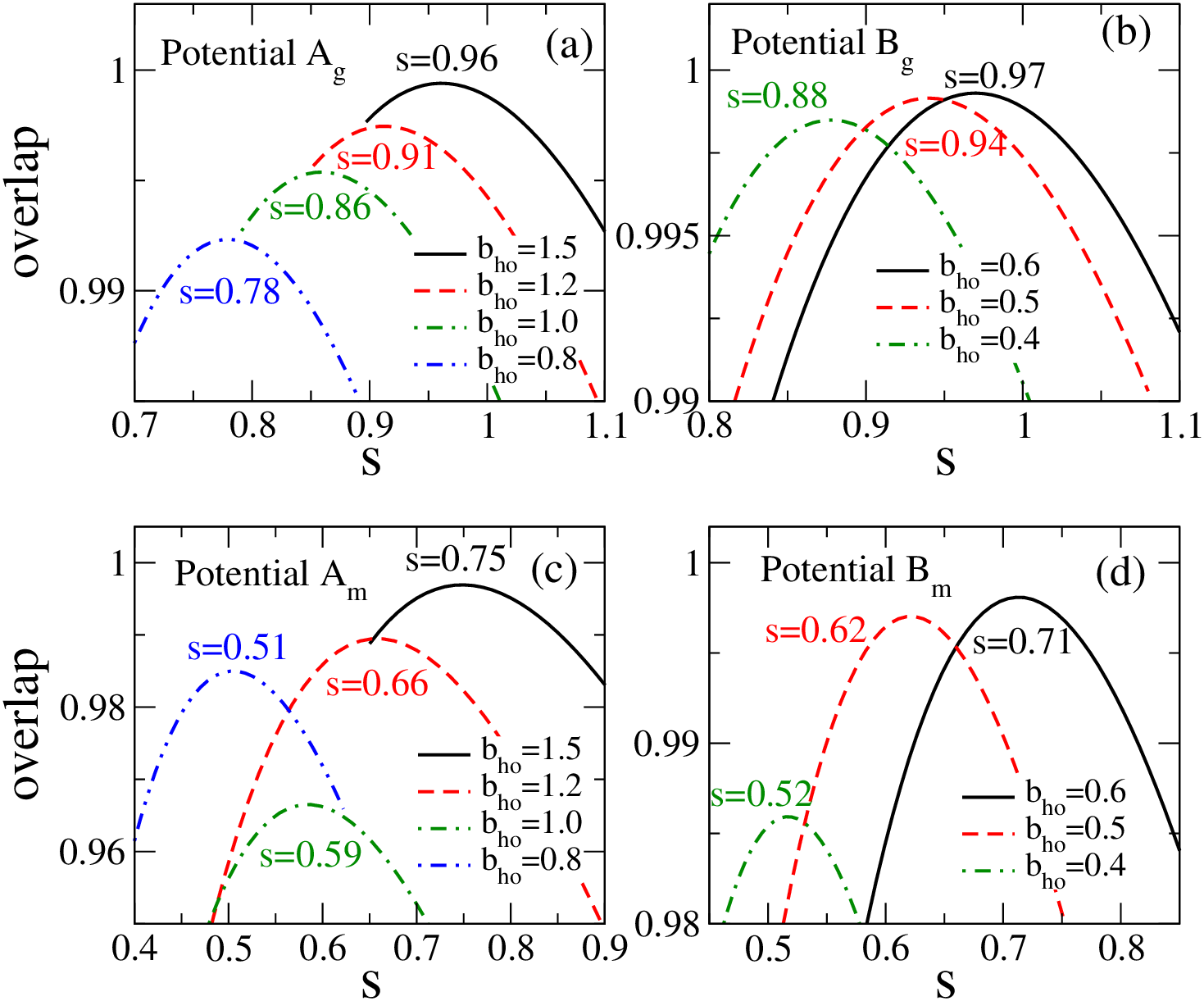}
\caption{Overlap ${\cal O}(s)$ as a function of the scale parameter $s$ for potentials A$_g$, B$_g$, A$_m$, and B$_m$ (panels (a), (b), (c), and (d), respectively) and different values of the external potential parameter $b_{ho}$. For each $b_{ho}$ the dimension $d$ is the one giving rise to the same ground state binding energy. For each case we indicate the value of $s$ corresponding to the maximum 
overlap.}
\label{fig6}
\end{figure}

In Fig.~\ref{fig6} we show the overlap ${\cal O}(s)$ as a function of the scale parameter, $s$, for the the
four potentials considered in this work. Several different values of the squeezing parameter, $b_{ho}$, have been
chosen, and for each of them we determine, according to Fig.~\ref{fig5}, the value of the dimension,
$d$, giving rise to the same ground state binding energy. As seen in Fig.~\ref{fig6}, in all the cases the
curves show a well defined maximum, which determine the value of the scale parameter, $s$, to be used
in the $d$-function $\tilde{\Psi}_d$.

\begin{table}
\begin{tabular}{|c|cc|cc|cc|cc|}\hline
$b_{ho}$  &\multicolumn{2}{c|}{Pot. A$_g$} &\multicolumn{2}{c|}{Pot. B$_g$} &\multicolumn{2}{c|}{Pot. A$_m$}&\multicolumn{2}{c|}{Pot. B$_m$} \\
   &   $s$ &  $d$ & $s$ &  $d$ & $s$ &  $d$ & $s$ &  $d$ \\ \hline
  0.4  &    &     &  0.88  & 2.489    &     &      &  0.52   & 2.205    \\
  0.5  &    &     &  0.94  &  2.604   &     &      &   0.62  &  2.318   \\
  0.6  &    &     &  0.97  &  2.694   &     &      &    0.71 &   2.383  \\
  0.8  & 0.78   &  2.523   &    &     & 0.51    &  2.173    &     &     \\
  1.0  &  0.86  &  2.628   &    &     &  0.59   &  2.289    &     &     \\
  1.2  &  0.91  &   2.704  &    &     &   0.66  &  2.400    &     &     \\
 1.5  &   0.96 &  2.783   &    &     &    0.75 &   2.511   &     &     \\ \hline
\end{tabular}
\caption{For each of the potentials and each of the selected values of $b_{ho}$ shown in Fig.\ref{fig6}, we give the corresponding 
values of the scale parameter $s$ that maximizes the overlap in Eq.(\ref{ovl}) and the dimension $d$ giving rise to the same 
ground state binding energy. }
\label{tab3}
\end{table}

Fig.~\ref{fig6}a shows the results for potential A$_g$. As we can see, for $b_{ho}=1.5$,
which corresponds to $d=2.783$, we obtain a scale parameter of $s=0.96$, which indicates that 
the deformation of the $d$-calculated wave function produced by the squeezing is in this case rather modest.
For larger squeezing (smaller values of $b_{ho}$), the scale
parameter starts decreasing, reaching the values of $s=0.91$, $s=0.86$, and $s=0.78$ for $b_{ho}=1.2$ ($d=2.704$),
$b_{ho}=1.0$ ($d=2.628)$, and $b_{ho}=0.8$ ($d=2.523$), respectively. These results are collected
in the second column of Table~\ref{tab3}.

The deformation of the $d$-calculated wave function produced by a given squeezing is very sensitive to the size of
the three-body system. For instance, as seen in Table~\ref{tab2}, for potential B$_g$ the ground state in 3D is
about three times smaller than the one for potential A$_g$. For this reason one can intuitively think that a larger
squeezing (smaller $b_{ho}$) will be necessary in order to get a similar deformation of the wave function. This is actually seen in Fig.~\ref{fig6}b,
where the overlap ${\cal O}(s)$ for potential B$_g$ is shown (the corresponding $s$ and $d$ values 
are given in the third column of Table~\ref{tab3}). As we can see, even for a squeezing parameter
$b_{ho}=0.6$ ($d=2.694$) the scale parameter is still not too far from 1, whereas for $b_{ho}=0.5$ ($d=2.604$)
and $b_{ho}=0.4$ ($d=2.489$) the corresponding values of $s$ are similar to the ones obtained for potential A$_g$ (Fig.\ref{fig6}a) 
when $b_{ho}=1.2$ and $b_{ho}=1.0$, respectively.

In Figs.~\ref{fig6}c and \ref{fig6}d we show the same results for the Morse-like potentials A$_m$ and B$_m$, respectively,
and the values of $s$ and $d$ are also given in the fourth and fifth columns of Table~\ref{tab3}.
The general behavior is similar to the one found for the Gaussian potentials. The three-body system is again about three
times bigger in 3D with potential A$_m$ than with potential B$_m$. The result is therefore that similar deformation, i.e. similar 
values of the scale parameter $s$, requires less squeezing, larger value of $b_{ho}$, with potential A$_m$ than with potential B$_m$.
The same happens when comparing potentials A$_m$ with A$_g$, and B$_m$ with B$_g$. Since the Morse 3D
states are clearly bigger than the corresponding Gaussian counterparts, we again find that for equal values of $b_{ho}$
the bigger systems, the ones obtained with potentials A$_m$ and B$_m$, are more deformed (smaller value of $s$)
than with potentials A$_g$ and B$_g$, respectively.

The maximum overlap shown in Fig.~\ref{fig6} is always very close to 1, always above 0.96, but typically around 0.98 
or even higher. This indicates that the deformation described by Eqs.(\ref{defx}) and (\ref{defy}) works in general very well,
although for some of the cases, like for instance potential A$_m$ with $b_{ho}=1.0$ (where the maximum overlap
is about 0.96), a correction could probably be introduced. 

\begin{figure}[t]
\includegraphics[width=0.95\linewidth]{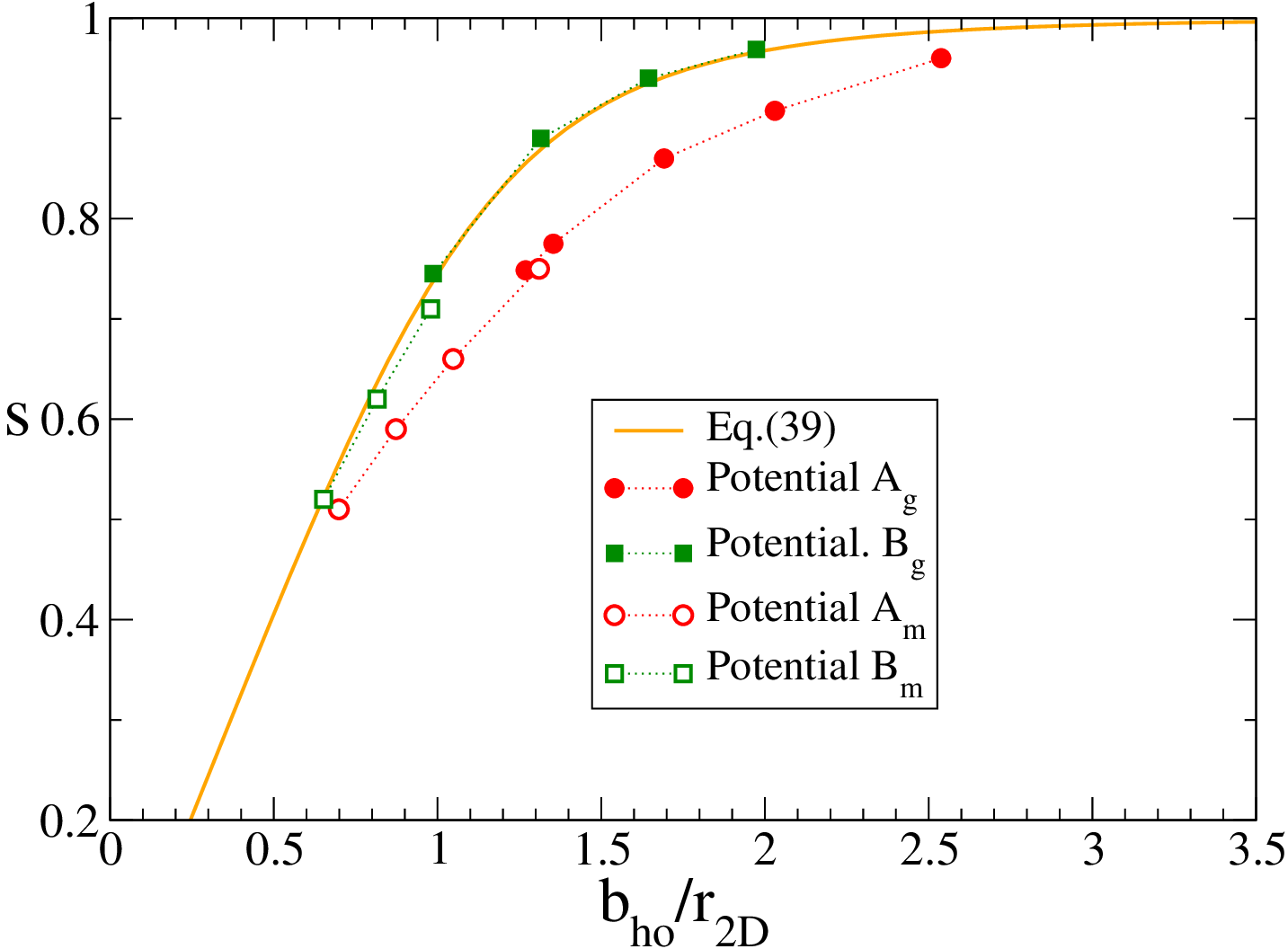}
\caption{Scale parameter $s$ as a function of $b_{ho}/r_{2D}$ for the four potentials used in this work. The solid
curve shows the estimate given in Eq.(\ref{estim}).}
\label{fig7}
\end{figure}

The values of the computed scale parameters are shown in Fig.~\ref{fig7} as a function
of $b_{ho}/r_{2D}$. Together with them we show the estimate given in Eq.(\ref{sbd}). More precisely, exploiting
again the fact that in the case of the harmonic oscillator two-body potential we have that $b_{pp}=r_{2D}\sqrt{3/2}$, we can then
rewrite Eq.(\ref{sbd}) as: 
\begin{equation}
\frac{1}{s^2} =\sqrt{ 1+\frac{9}{4}  \left(\frac{r_{2D}}{b_{ho}}\right)^4 },
\label{estim}
\end{equation}
which is shown in Fig.~\ref{fig7} by the solid curve. As we can
see, the estimate given above works very well for potentials B$_g$ and B$_m$ (squares in the figure), which correspond
to the cases of well bound three-body states in 3D (Table~\ref{tab2}). In fact, for potential 
B$_g$, for which $|E_{3D}|$ is pretty large, the agreement is excellent. For the cases
when the system is clearly less bound in 3D, potentials A$_g$ and A$_m$ (circles in the figure), the computed results
disagree with the estimate in Eq.(\ref{estim}) in the region of intermediate squeezing. However, it is
interesting to see how, for large squeezing, the computed curve for potentials A$_g$ and A$_m$  and
the solid curve clearly converge. Therefore the estimate (\ref{estim}) appears as a very good way of determining the
equivalence between $s$ and $b_{ho}$ also for potentials A$_g$ and A$_m$ in the cases of large squeezing, which,
on the other hand, are the cases where the calculations with the external potential are more problematic.

Finally, let us directly compare the wave functions $\Psi_{ext}$ and $\tilde{\Psi}_d$. This can be done
by simple comparison of the radial wave functions contained in the expansions Eqs.(\ref{wf0}) and (\ref{wf1}). Since
the angular functions are the same in the two cases, the difference between the two wave functions will be
necessarily contained in the radial functions $f_n(\rho)$ and $\tilde{f}_n^{(d)}(\rho,s)$.

\begin{figure}[t]
\includegraphics[width=0.95\linewidth]{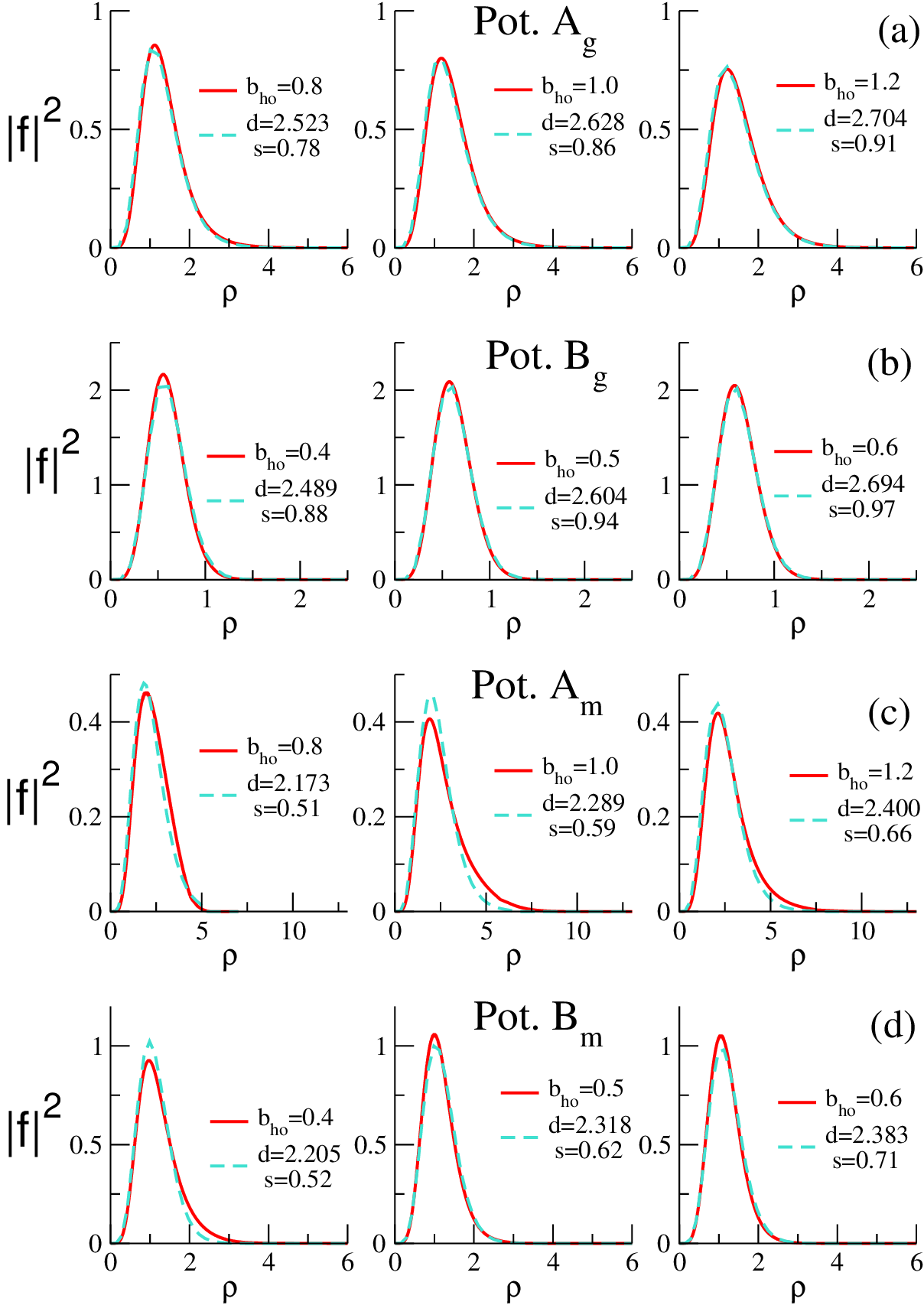}
\caption{The solid curves show, as a function of $\rho$, the square of the dominating radial wave functions, $f_n(\rho)$ 
in Eq.(\ref{wf0}), for the potentials and external field parameters $b_{ho}$ shown in Fig.~\ref{fig6} (except the cases
with $b_{ho}=1.5$, which, for simplicity in the figure, have been omitted). Panels
(a), (b), (c), and (d) correspond to potentials A$_g$, B$_g$, A$_m$, and B$_m$, respectively. For each case, the dashed curve gives,
also as a function of $\rho$, the square of the radial wave function $\tilde{f}_n^{(d)}(\rho,s)$ shown in Eq.(\ref{rad1}), where
the dimension $d$ provides the same ground state energy as the calculation with the external field, and the scale parameter $s$, given
in Fig.~\ref{fig6}, maximizes the overlap ${\cal O}(s)$ in Eq.(\ref{ovl}). The length unit is the range of the 
two-body interaction.}
\label{fig8}
\end{figure}

In Fig.~\ref{fig8} we show the squares of the dominating radial wave functions,
$f_n(\rho)$ (solid curves) and $\tilde{f}_n^{(d)}(\rho,s)$ (dashed curves),
for some of  the cases shown in Fig.~\ref{fig6} (in order to make the figure simpler
the cases with $b_{ho}=1.5$ have been omitted).  In all of them, the dominating term provides
 more than 98\% of the total norm. Panels (a), (b),(c), and (d) correspond to the results with potentials 
A$_g$, B$_g$, A$_m$, and B$_m$, respectively. The values of the dimension, $d$, and the scale
 parameter, $s$, are as indicated in Table~\ref{tab3}, i.e. the dimension, $d$, is such that the $d$-calculation
 gives the same ground state energy as in the calculation with the external squeezing potential, 
 and the scale parameter, $s$, is such that the overlap, ${\cal O}(s)$, in
 Eq.(\ref{ovl}), is maximum. 
  
 As we can see in the figure the agreement is very good basically for all the cases shown. The largest
discrepancy is observed for Morse potential A$_m$ with $b_{ho}=1.0$, which, as seen in Fig.~\ref{fig6},  
is the case with the smallest value of the maximum overlap, below 0.97, between the two wave functions. In any
case, the good agreement between the two radial wave functions illustrates how the wave function resulting
from the $d$-calculation, after the appropriate reinterpretation in the ordinary 3D space, provides
a very good description of the system squeezed by an external potential.

\section{Summary and conclusions}
\label{sec10}

In this work the continuous squeezing of three-body systems from 3 to 2 dimensions
has been investigated by means of two different procedures: First, a method
where the external confining potential is explicitly included, and second, a method
where the external potential does not enter, but instead, the dimension $d$ is allowed
to take any intermediate value between $d=3$ and $d=2$.

The case of three identical spinless bosons with relative $s$-waves between all the particles
is considered. For the two-body potentials we have used Gaussian and Morse radial forms whose scattering length
 vary from small, with a few fairly well bound states, to rather large.

For the three-body calculations with an external field we have chosen the
squeezing potential to be a one-body deformed harmonic oscillator acting on
each of the three particles.  The oscillator length in one of the dimensions
is varied from infinity (no squeezing) to a very small value
corresponding to a fraction of the two-body interaction range.
The deformation produced by the squeezing
breaks orbital angular momentum conservation, and the calculations must
account for the resulting mixing of these quantum numbers.
The consequence is that  the
adiabatic hyperspherical expansion must employ a large number of
partial wave components in the limit where the dimension $2$ is
approached.

The method with the non-integer dimension, $d$, as parameter is
formulated in terms of hyperspherical coordinates with spherical
potentials and $s$-waves. The phase space and centrifugal
barrier potentials both correspond to the specified value of $d$. 
This method is technically precisely as
simple as the ordinary three-body three-dimensional computations
without external field.  

With the two methods we have investigated how the ground state 
energies vary when squeezing from three to two dimensions. Both
methods show the same qualitative behavior although the connection
between the 2D and 3D limits must be different in the two cases, since
the oscillator length of the external potential and the dimension, $d$,
vary in infinite and finite intervals, respectively. The three-body bound states 
do not necessarily have counterparts in the 3D and the 2D limits.

Comparison of the results from the two methods allows extraction of
the function translating between $d$ and the oscillator length of the
external squeezing potential.  The knowledge of this function 
provides a subsequent prediction of which external
field corresponds to a given $d$-value.  To give us some insight, we have used
a simplified system where harmonic oscillators (different to the external
field) are used for the two-body interactions. We have in this way obtained an analytic
expression relating the dimension $d$ and the oscillator length of the
external field. This estimate is then compared to the results arising from the numerical
computations.
By choosing the root-mean-square radius of the three-body system in two
dimensions as length unit, we have found that the translational functions are
remarkably similar to each other, as well as to
the analytic function derived for the two-body oscillator potential.

To be able to predict any observable property entirely from $d$-calculations, we
must also find a translation or interpretation providing the wave function
corresponding to a three dimensional calculation with an external
field.  We naturally search for a deformed solution obtained from the
spherical wave function in the $d$-method.  This is achieved by
scaling the squeezing coordinate relative to the two other
coordinates. The scaling factor must depend on $d$, and it is defined such
that the overlap between the two wave functions of equal energy from 
the two methods is maximum, preferentially unity.  

This interpretation of the $d$-calculated wave function is suggested by the exact
solution available when using oscillator interactions between the particles as
well as for the external field. This analytic approximation
 is compared to the scaling parameter extracted numerically. The computed values are surprisingly close
 to the analytic curve for the potentials with well bound three-body states in 3D, whereas
 they deviate when the three-body system is weakly bound in three dimensions.
 Comparing directly the wave functions from the two methods  we find a remarkable agreement
 between them.

In summary, we have in details investigated a new method to deal with
three-body problems in non-integer dimensions between $2$ and $3$.
The basis is a $d$-dimensional phase space, a corresponding
$d$-dependent generalized centrifugal barrier, and a spherical
computation.  The calculations are precisely as simple as the same
three-body calculations in integer dimensions.  We validate the method
by comparing to the brute force method in the ordinary three-dimensional
space.  The non-integer dimension is simulated by a deformed external
field, which effectively reduces movement in one coordinate from being
free (in 3D) to zero space (in 2D).

The equivalence of the two methods is shown for identical bosons by
numerical calculations, where we provide a relatively accurate
translation between the parameter $d$ and the external field length.
This connection includes a prescription to obtain the complete
deformed wave function from the $d$-calculation.  Since the
translation correctly reproduces the two limits of 2D and 3D, any
prediction occurring for some $d$-value is bound to happen for some
external field strength, which in turn is relatively precisely given
by an analytic expression.

The presented $d$-method can be extended to asymmetric three-body systems with
different mass ratios, to dimensions smaller than 2 and larger than 3,
and perhaps to fermions and particle numbers larger than three.

\appendix
\section{$s$-wave spherical and hyperspherical harmonics in $d$ dimensions}
\label{app}

The angle independent $s$-wave spherical harmonic in $d$ dimensions, $Y_0$, 
can be  obtained by taking into account that the phase volume in $d$
dimensions is given by \cite{hay01}: 
\begin{equation}
\int d\Omega_d=\frac{2\pi^{d/2}}{\Gamma\left(\frac{d}{2}\right)}, 
\end{equation}
from which, since $\int Y_0^* Y_0 d\Omega_d=1$, one immediately gets:
\begin{equation}
Y_0=\left[ \frac{\Gamma\left(\frac{d}{2}\right)}{2\pi^{d/2}}\right]^{1/2}.
\end{equation}

Thus, the angular functions, $\Phi_n^{(d,i)}$, in Eq.(\ref{eq4}) depend only on 
the hyperangle $\alpha=\arctan(x/y)$, where $\bm{x}$ and $\bm{y}$ are
the usual Jacobi coordinates \cite{nie01}. Typically, these angular functions are obtained as an expansion 
in terms of the hyperspherical harmonics, which for $d$ dimensions and
relative $s$-waves, take the form \cite{nie01}:
\begin{equation}
{\cal Y}_K^{(d)}(\alpha)=N_\nu P_\nu^{\left(\frac{d-2}{2},\frac{d-2}{2}\right)}(\cos 2\alpha) Y_0 Y_0,
\label{hh}
\end{equation}
where $K=2\nu$ is the hypermomentum, $N_\nu$ is the normalization constant, and $P_\nu^{\left(\frac{d-2}{2},\frac{d-2}{2}\right)}$ is a Jacobi polynomial.

Finally, when solving the three-body problem it is always necessary at some point
to rotate the hyperspherical harmonics (\ref{hh}) from one Jacobi set $i$ into a different set $j$.
This is done by means of the Raynal-Revai coefficients, which for $s$-waves and $d$ dimensions
are given by \cite{nie01}:
\begin{equation}
\langle {\cal Y}_K(i) |  {\cal Y}_{K'}(j)\rangle = \delta_{KK'}
\frac{P_\nu^{\left(\frac{d-2}{2},\frac{d-2}{2}\right)}(\cos 2\gamma_{ij})}
                      {P_\nu^{\left(\frac{d-2}{2},\frac{d-2}{2}\right)}(1)},
\end{equation}
where the angle $\gamma_{ij}$ is given for instance in Ref.\cite{nie01}.

\end{document}